\documentclass[prb,showpacs,twocolumn,superscriptaddress,floatfix,longbibliography]{revtex4-2}
\usepackage{graphicx,amsfonts,amssymb,amsthm,amsmath,xspace,verbatim}
\usepackage[colorlinks=true,citecolor=green,linkcolor=red,urlcolor=blue]{hyperref}
\usepackage[T1]{fontenc}
\usepackage{subfigure}
\usepackage{blindtext}

\usepackage{color}
\definecolor{g}{rgb}{.1,0.4,.1} 
\definecolor{b}{rgb}{0,0.2,1}
\definecolor{rouge}{rgb}{0.82,0.,0.}
\definecolor{vert}{rgb}{0.,0.82,0.}
\definecolor{orange}{rgb}{1,0.5,0.}
\definecolor{bleu}{rgb}{0.,0.,0.82}
\definecolor{m}{rgb}{0.82,0.,0.82}
\definecolor{vert2}{rgb}{0.,0.5,0.}
\definecolor{rougeclair}{rgb}{1.0,0.7,0.7}

\newcommand{\beq}{\begin{equation}}
\newcommand{\be}{\begin{equation}}
\newcommand{\beqn}{\begin{eqnarray}}
\newcommand{\eeq}{\end{equation}}
\newcommand{\ee}{\end{equation}}
\newcommand{\eeqn}{\end{eqnarray}}

\newcommand{\bem}{\begin{pmatrix}}
\newcommand{\eem}{\end{pmatrix}}

\newlength{\ldag}
\begin{document}

\title{Aharonov-Bohm cages, flat bands, and gap labeling in hyperbolic tilings}

\author{R\'emy Mosseri}
\email{remy.mosseri@upmc.fr}
\affiliation{Sorbonne Universit\'e, CNRS, Laboratoire de Physique Th\'eorique de la Mati\`ere Condens\'ee, LPTMC, 75005 Paris, France}

\author{Roger Vogeler}
\email{vogelerrov@ccsu.edu}
\affiliation{Department of Mathematical Sciences, Central Connecticut State University, New Britain, Connecticut 06050, USA}

\author{Julien Vidal}
\email{vidal@lptmc.jussieu.fr}
\affiliation{Sorbonne Universit\'e, CNRS, Laboratoire de Physique Th\'eorique de la Mati\`ere Condens\'ee, LPTMC, 75005 Paris, France}

\begin{abstract}

Aharonov-Bohm caging is a localization mechanism stemming from the competition between the geometry and the magnetic field. Originally described for a tight-binding model in the dice lattice, this destructive interference phenomenon prevents any wavepacket spreading away from a strictly confined region. Accordingly, for the peculiar values of the field responsible for this effect, the energy spectrum consists of a discrete set of highly degenerate flat bands. In the present work, we show that  Aharonov-Bohm cages are also found in an infinite set of hyperbolic dice tilings defined on a negatively curved hyperbolic plane.
We detail the construction of these tilings and compute their Hofstadter butterflies by considering periodic boundary conditions 
on high-genus surfaces. As recently observed for some regular hyperbolic tilings, these butterflies do not manifest the self-similar structure of their Euclidean counterparts but still contain some gaps. 
We also consider the energy spectrum of  hyperbolic kagome tilings (which are the dual of hyperbolic dice tilings), which displays interesting features, such as highly degenerate states arising for some particular values of the magnetic field. For these two families of hyperbolic tilings, we compute the Chern number in the main gaps of the Hofstadter butterfly  and propose a  gap labeling inspired by the Euclidean case. Finally, we also study the triangular Husimi cactus, which is a limiting case in the family of hyperbolic kagome tilings, and we derive an exact expression for its spectrum versus magnetic flux.

\end{abstract}

\maketitle


%
%
\section{Introduction } 
%
%

Two-dimensional electron systems in a perpendicular magnetic field have been intensively studied in condensed-matter physics, in particular in the context of integer~\cite{Klitzing80} and fractional~\cite{Tsui82} quantum Hall states or the self-similar Hofstadter butterfly~\cite{Hofstadter76} describing the structure of energy levels for tight-binding electrons versus the magnetic flux.
This subject is not limited to condensed-matter experiments, as proved more recently in the field of atomic physics with the possibility of generating ``artificial" magnetic fields acting on cold-atom assemblies, which opens the way to different types of experiments in that direction~\cite{Bloch12}.

In 1998, an extreme localization effect induced by the magnetic field was discovered, for tight-binding models in certain periodic tilings, such as the dice tiling~\cite{Vidal98} or the diamond chain~\cite{Vidal00}, at a critical value $f_c=1/2$ of the magnetic flux per plaquette (measured in units of the quantum flux $\phi_0$). Aharonov-Bohm (AB) cages are shown to exist due to a complete destructive interference affecting  a particle's motion. These cages have a spectral signature: the dice tiling butterfly displays a density of states that pinches while approaching $f_c$, leading at $f_c$ to an energy spectrum consisting of three highly degenerate energy levels.

Among these three energy levels, the one at zero energy is present at any flux, and the other two are the result of destructive interferences tuned by the magnetic field. A particle, initially located on a site of the tiling, displays a quantum diffusion limited to a small cluster of sites (the so-called AB cage), and eventually  bounces back and forth to its original position. This effect disappears if the flux is tuned away from $f_{\rm c}$~\cite{Vidal98} or if interactions between particles~\cite{Vidal00} or disorder~\cite{Vidal01} are introduced. Notice that AB cages are not limited to tight-binding systems, and were recently shown to occur for quantum walks~\cite{Per20,Per22}.

These AB cages have triggered much interest and have been observed in different experimental setups such as superconducting wire networks~\cite{Abilio99}, Josephson junction arrays~\cite{Pop08}, cold atomic gases~\cite{Moller12}, photonic lattices~\cite{Mukherjee18}, ion micro traps~\cite{Bermudez11}, and more. For a recent review concerning artificial systems, see Ref.~\cite{Flach18}.

So far, Hofstadter butterflies have been mainly investigated for Euclidean tilings, either periodic~\cite{Hofstadter76,Claro79} or quasiperiodic~\cite{Vidal04,Fuchs16}. The case of regular tilings of the hyperbolic plane $\mathbf{H}^2$  has also been recently addressed~\cite{Stegmaier22}. In this case, Hofstadter  butterflies display no self-similarity and only a few gaps. Regular two-dimensional hyperbolic  tilings are well known and classified in standard mathematical texts~\cite{Magnus74}. Their underlying negatively curved geometry was  considered in the context of geometric frustration~\cite{Kleman79,Rubinstein03,Sausset07}, and the fact that limiting cases of these tilings correspond to embeddings of the Bethe lattice and the Husimi cactus was noticed~\cite{Mosseri82}. Recently, there has been a renewed interest in these tilings in the context of quantum information~\cite{Breuckmann16}, of circuit quantum electrodynamics~\cite{Kollar19,Kollar20} and of electrical circuits~\cite{Lenggenhager22}.

Here, we study two infinite families of semiregular hyperbolic tilings. For the first one (which generalizes the dice tiling Euclidean case), we show the occurrence of AB cages. The second family, dual to the first one, is a generalization to $\mathbf{H}^2$ of the standard (Euclidean) kagome tiling.  This case also displays interesting features, like a spectral pinching phenomenon on a simple flux-dependent curve, and the existence of many gaps.

The article is organized as follows. In Sec.~\ref{sec:tilings}, we introduce the set of hyperbolic dice and kagome tilings. We then study in Sec.~\ref{sec:butterfly} their associated Hofstadter butterflies, and analyze the triangular Husimi cactus case. Section~\ref{sec:gaplabel} is devoted to the analysis of the gap-labeling problem.  Several Appendices provide additional information.

%
%
\section{Hyperbolic dice and kagome tilings}
\label{sec:tilings}
%
%
\subsection{Regular two-dimensional tilings}
%
%
The Schl\"afli symbol $\{p,q\}$ is a standard notation for regular tilings made of a regular polygon with $p$ sides (a $p$-gon), such that each site is shared by $q$ $p$-gons~\cite{coxeter73}. To a given $\{p,q\}$, one can associate a dual tiling $\{q,p\}$, sharing the same symmetry group, whose vertices are located at the center of the $p$-gons of the $\{p,q\}$ tiling. These tilings are compatible with one of the three two-dimensional geometries with constant Gaussian curvature:

%
%
\begin{enumerate}
\item  $(p-2)(q-2)<4$ denotes the five Platonic polyhedra---the  self-dual tetrahedron $\{3,3\}$, the dual  $\{3,4\}$ octahedron and $\{4,3\}$ cube, and the dual $\{3,5\}$ icosahedron  and $\{5,3\}$ dodecahedron---which can be embedded in the positively curved sphere $\mathbf{S}^2$.
\item $(p-2)(q-2)=4$ denotes the self-dual $\{4,4\}$ square tiling, and the dual $\{3,6\}$ triangular and $\{6,3\}$ hexagonal tilings which live on the  Euclidean (flat) plane.
\item $(p-2)(q-2)>4$ denotes an infinite set of tilings of the negatively curved hyperbolic plane $\mathbf{H}^2$.
\end{enumerate}
%
%

The symmetry group of a $\{p,q\}$ tiling, denoted $[p,q]$, is generated by reflections in the sides of a characteristic (also called ``orthoscheme") triangle with angles $\pi/p$, $\pi/q$ and $\pi/2$ (see  Fig.~\ref{fig:orthoscheme37} for the $\{7,3\}$ hyperbolic case).

This apparent hyperbolic space richness (with an infinite set of regular tilings as compared to the finite sets found in  spherical and Euclidean spaces) is a peculiar two-dimensional property. Indeed, the number of regular tessellations in higher-dimensional hyperbolic spaces is finite and small.

%
%
\subsection{Hyperbolic tilings}
\label{tilings}
%
%

In this work, we focus on semiregular hyperbolic tilings related to the infinite set of triangular tilings $\{p,3\}$. To represent them, we use the Poincar\'e disk conformal representation, such that $\mathbf{H}^2$ points are located inside a unit disk (whose boundary is the locus of points at infinity). In this representation, $\mathbf{H}^2$ geodesic lines are circular arcs orthogonal to the unit circle. Reflections about these geodesics are inversions with respect to these circles~\cite{Magnus74,Boettcher22}.

\begin{figure}[t]
\centering
\includegraphics[width=0.5\textwidth]{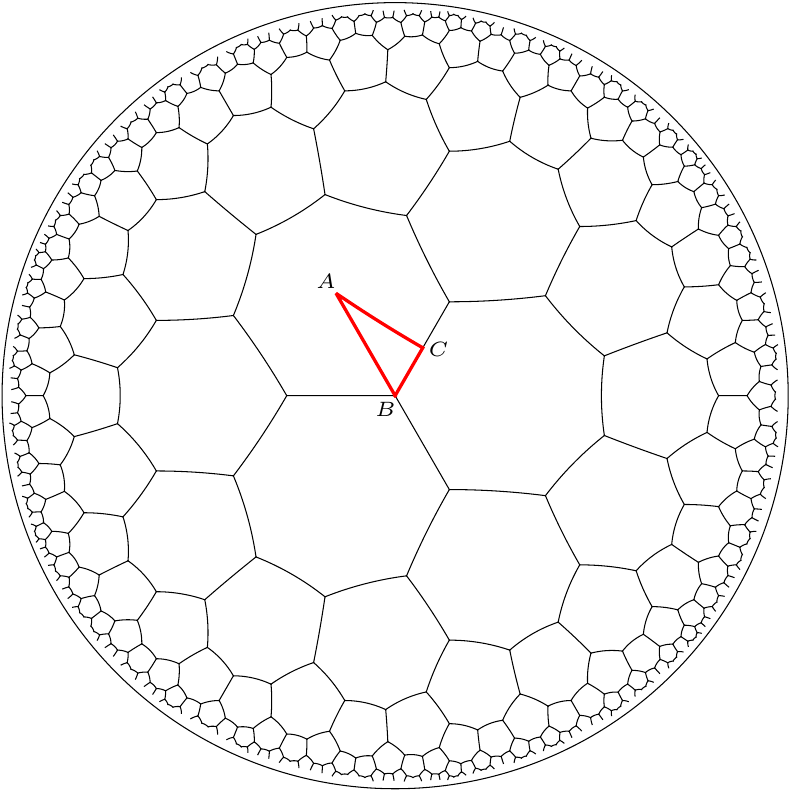}
\caption{A piece of the $\{7,3\}$ hyperbolic tiling together with the orthoscheme triangle ($A,B,C$), with red edges. The symmetry group $[3,7]$ is generated by (hyperbolic) reflections with respect to the sides of this triangle. The resulting images of vertices $A$, $B$, and $C$ lead respectively  to the $\{3,7\}$, $\{7,3\}$, and the kagome $\mathcal{K}_7$ tilings.}\label{fig:orthoscheme37}
\end{figure}

 Figure~\ref{fig:orthoscheme37} shows the orthoscheme triangle associated to the hyperbolic tiling $\{7,3\}$. The latter is the image of site $B$ under reflections in the sides of the orthoscheme triangle, while $\{3,7\}$ is the image of site $A$. Equivalent constructions can be made for larger and larger polygons, up to polygons of infinite size, called apeirogons or $\{\infty\}$. The latter lead to regular $\{ \infty ,q\}$  tilings which are explicit realizations of infinite regular treelike structures, called Bethe lattices~\cite{Mosseri82}. It is well known that a regular tree cannot be isometrically embedded in a Euclidean plane without self-crossing. This becomes possible in $\mathbf{H}^2$ owing to the fact that for an $\mathbf{H}^2$ disk, the boundary grows exponentially with the radius.
 
%
%
%
%

%
%
\subsection{From regular $\{p,3\}$ to hyperbolic dice and kagome tilings}
%
%

\begin{figure}[t]
\includegraphics[width=0.5\textwidth]{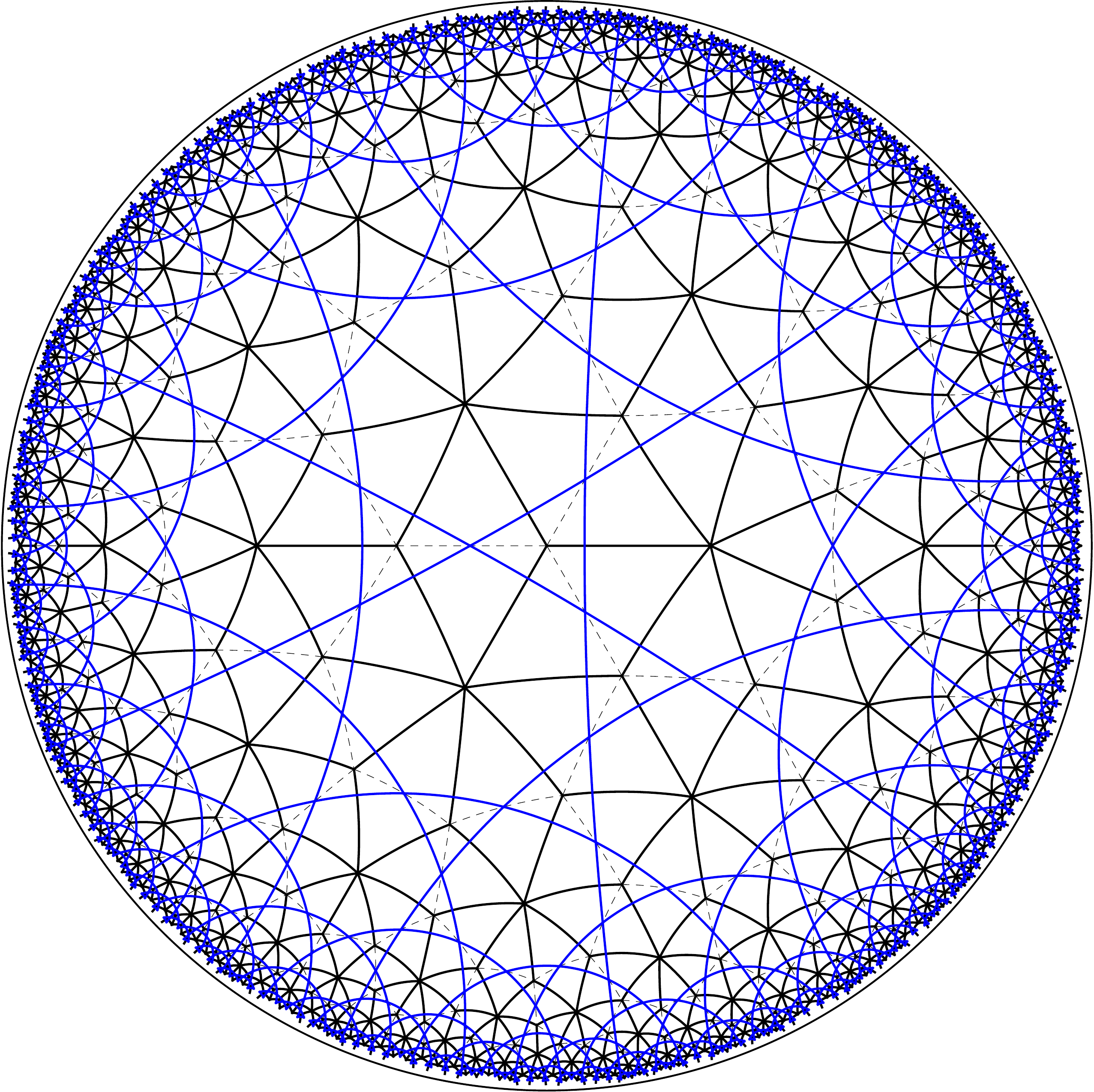}
\caption{A piece of the hyperbolic dice $\mathcal{D}_7$ tiling (with black edges forming rhombi) and its dual kagome $\mathcal{K}_7$ tiling (with blue edges forming triangles and heptagons). The associated $\{7,3\}$ tiling edges are shown as dashed lines.}
\label{fig:dicekagome}
\end{figure}

We first describe an infinite set of semiregular rhombus tilings derived from $\{p,3\}$ tilings. They have threefold and $p$-fold coordinated sites as vertices, generalizing the Euclidean dice tiling, and we denote them by $\mathcal{D}_p$. The vertex set of $\mathcal{D}_p$ is simply the union of all the sites of $\{3,p\}$ with all the sites of $\{p,3\}$. The edge graph is bipartite, with each edge connecting a $\{p,3\}$ site to a $\{3,p\}$ site. The rhombic faces are all congruent, with area equal to $2/3$ of the $\{3,p\}$ triangle area. An example is illustrated in Fig.~\ref{fig:dicekagome} for $p=7$.

We  also consider the tilings dual to $\mathcal{D}_p$. These are generalized  kagome tilings, which we denote here by $\mathcal{K}_p$ instead of the more conventional symbol $\left\{ \begin{matrix} p \\ q \end{matrix}\right\} $~\cite{coxeter73}.   The faces of $\mathcal{K}_p$ are of two types, triangles and $p$-gons.  Its sites are all alike, and located at the midpoints of the  $\{p,3\}$ edges (see point $C$ in Fig.~\ref{fig:orthoscheme37}), which are also midpoints of the $\{3,p\}$ edges.  In graph theory, the edge graph of  $\mathcal{K}_p$ is called the line graph of the $\{p,3\}$ edge graph~\citep{Biggs74}(see again Fig.~\ref{fig:dicekagome} for $p=7$).

%
%
 %
 %

%
%
\subsection{Boundary conditions}
%
%

Our main results will concern tight-binding spectra under a perpendicular magnetic field (see Sec.~\ref{sec:butterfly}).   Regular Euclidean tilings are periodic, which allows one, in zero field, to use Bloch's theorem. Switching on the magnetic field already adds  difficulties in the Euclidean case, since the gauge that enters the Hamiltonian generically breaks tiling symmetries~\cite{Hofstadter76}.

The situation is highly more complex in the hyperbolic case, since the tiling symmetry group $[p,q]$ is non-Abelian, which in principle prohibits the application of Bloch theory, even at zero magnetic flux. Note, however, a recent proposition~\citep{Maciejko21} to build some (Bloch-like) eigenvectors for hyperbolic tilings, using one-dimensional irreducible representations (irreps) of a (still non-Abelian) Fuchsian subgroup of $[p,q]$ generated by translations on $\mathbf{H}^2$. These symmetry group elements, as for standard Euclidean translations, are fixed-point free, and can generate the whole tiling by repeated action on an associated unit cell. It is not yet known which part of the spectrum can be captured by these Bloch-like states.
 
To compute spectral properties associated with the above-described hyperbolic tilings, we need to specify the boundary conditions. Two main such conditions could be applied, which are now discussed.

%
%
\subsubsection{Open boundary conditions}
%
%
Let us first discuss open boundary conditions (OBCs). A finite patch of the tiling has cut edges at its boundary, and this irregularity modifies the spectrum. In the Euclidean case the ratio of boundary sites to bulk  sites vanishes at the thermodynamic limit, and one expects that this limit is numerically reached by increasing the size of the tiling patches, and eventually proceeding to finite-size scaling analysis when needed. As said above, in the hyperbolic case, the ratio of boundary to bulk sites remains finite with increasing patch size, which makes OBCs quite unsuitable to use. Dealing with regular tilings, we could concentrate on the local eigenspectrum near the center of the patch, which should be closer to that expected for the infinite tiling. As is well known, the moments of the tight-binding local density of states (LDOS) are proportional to the number of closed paths starting at the considered site. Density-of-states moments are therefore correct up to the graph diameter of the OBC patch.
 
%
%
\subsubsection{Periodic boundary conditions} 
\label{sec:PBC}
%
%
We now consider periodic boundary conditions (PBCs), and the way to approach the thermodynamic limit upon increasing the patch size. There are two main differences between the Euclidean and hyperbolic cases, in terms of topology and finite-size effects.
 
For Euclidean two-dimensional tilings, implementing PBCs amounts to mapping the tiling onto a (genus-1) torus, by identifying sites on opposite edges of a parallelogram patch. The main advantage is to suppress dangling edges on the boundary. But this, however, changes the LDOS whenever moments involve  noncontractible closed paths on the torus. All parallelograms, no matter how large, can be mapped to a genus-1 torus as long as the opposite sides are separated by (possibly large) translation symmetries of the full tiling. At vanishing magnetic flux, we can again apply Bloch theory and get the compact patch spectrum by selecting a discrete set of $k$-vectors in the Brillouin zone. The thermodynamic limit is smoothly approached by taking finer $k$-vector meshes, corresponding therefore to larger (compactified) patches of the tiling.  The smallest paths that are counted on a torus, and do not exist in the infinite structure are the noncontractible paths whose typical length is the diameter of the patch that is closed onto the torus, which goes as $\sqrt{V}$ for a patch with $V$ vertices. As a result, the spectrum will show higher LDOS moments, of order $\sqrt{V}$ and larger, as compared to the spectrum  of the infinite-size tiling.

The situation is very different in the hyperbolic case, since tiling  patches of increasing size will map onto tori of increasing genus, and therefore different topologies. This is due to the following well-known facts. For any tiling of a closed surface $M$ of genus $g$, the Euler-Poincar\'e formula relates the number of vertices $V$, edges $E$, and faces $F$ of the tiling to the Euler characteristic $ \chi =2 -2 g$ of $M$. In addition, the Gauss-Bonnet formula relates $\chi$ to the integral of the  Gaussian curvature $\kappa$ over $M$:
%
%
\beq
V-E+F = \chi = \frac {1}{2\pi} \iint\limits_M \kappa\, d\sigma.
\label{gaussbonnet}
\eeq
%
%

For a  hyperbolic tiling of type $\{p,q\}$, each $p$-gonal face carries the same total negative curvature, and the right-hand side of the above relation grows linearly (in absolute value) with $F$, implying that the genus increases with the size of the patch. We can go further and write, for several tiling families, simple relations between the genus and the number of sites:
%
%
\begin{align} \label{eq:nbsites}
\{p,3\} \rm{:} \quad &  V=4 p (g-1)/(p-6), \\
\{3,p\} \rm{:} \quad &  V=12 (g-1)/(p-6),   \\
\mathcal{D}_p \rm{:} \quad &  V=(4p+12) (g-1)/(p-6),\\
\mathcal{K}_p \rm{:} \quad &  V= 6p(g-1)/(p-6). 
\end{align}
%
%

This does not mean that any regular patch satisfying these relations can be mapped coherently onto a closed surface.  A nice family of highly symmetrical compact solutions, used in the present work, is provided by the set of $\{7,3\}$ Hurwitz tilings, described in Appendix~\ref{appendix:hurwitz}. Other examples, which include other  values of $p$, are obtained from the complete list of vertex-transitive threefold coordinated graphs  with up to 2048 vertices~\cite{Conder06}. This list contains, among others, many interesting $\{p,3\}$ tilings defined on $g$-holed tori, whose associated quantum properties under a perpendicular magnetic field have recently been studied~\cite{Stegmaier22}. 

Notice that the hyperbolic case is different from the Euclidean one in terms of finite-size analysis. Indeed, on a high-genus surface there will be short, noncontractible graph loops with lengths growing no better than linearly with the diameter of the patch, while $V$ grows exponentially with this diameter. As a consequence the finite-size difference in the computed spectrum  is shifted towards LDOS moments of order $\log{V}$ for a patch of $V$ vertices, meaning that the thermodynamic regime is quite slowly approached.
%
%
\begin{figure*}[t]
\includegraphics[width=\textwidth]{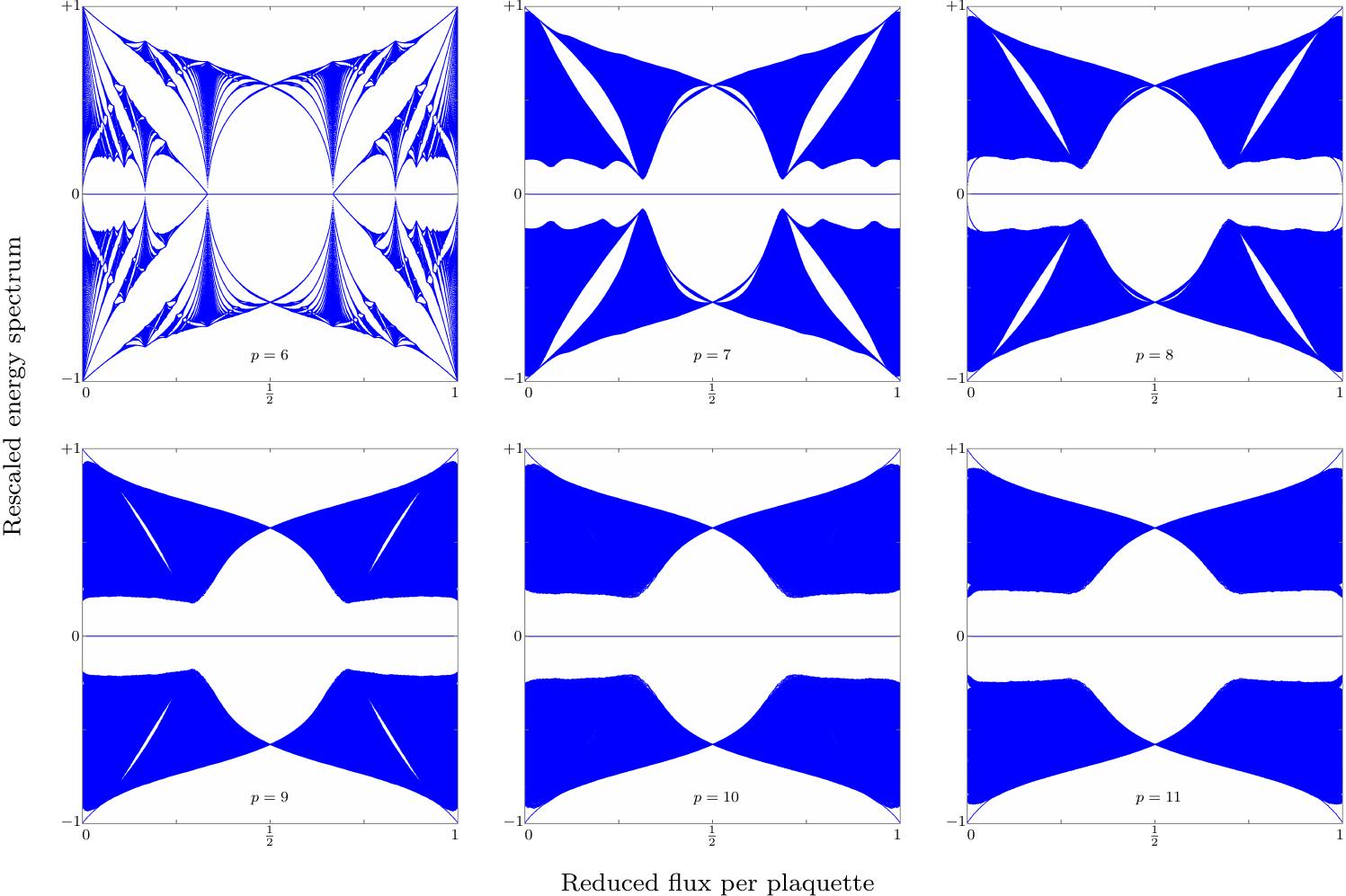}
\caption{Hofstadter butterflies for several dice $\mathcal{D}_p$ tilings, displaying the spectral support versus magnetic flux. More precisely the different structures are characterized by the triplet $(p$, symmetry; $g$ surface genus, $V$, number of $\mathcal{D}_p$ vertices) which reads here $(6, 1, 3024), (7, 118, 4680), (8, 129, 2816), (9, 169, 2688), (10, 169, 2184),  (11, 231, 2576)$. For $p=7$, the tiling is a member of the Hurwitz tiling series.
 All butterflies display the characteristic spectral shrinking near $f=1/2$, and a degenerate vanishing energy level for any flux. For the sake of comparison, energy spectra have been rescaled by a factor $\sqrt{3 p}$.}
\label{fig:Butterfly_dice_main}
\end{figure*}
%
%

%
%
\section{Hofstadter butterflies for hyperbolic tilings} 
\label{sec:butterfly} 
%
%
 %
%
\subsection{Hofstadter model} 
\label{sec:Hofstadter}
%
%
We consider the following tight-binding Hamiltonian: 
%
\begin{equation}
H=-\sum_{\langle i,j\rangle}t_{i,j} | i \rangle  \langle j |,
\label{eq:hamiltonian}
\end{equation}
%
%
%
%
where $\langle i,j\rangle$ stands for nearest-neighbor sites  and $t_{i,j}$ is the hopping amplitude. Without loss of generality, we set $t_{i,j}=1$ in the following. The effect of an external uniform magnetic field $\mathbf{B}$, perpendicular to the tiling, is taken into account by a modification of the hopping amplitude by a Peierls phase term~\cite{Peierls33}: 
%
%
\begin{equation}
t_{i,j}\rightarrow t_{i,j}\ \, {\rm e}^{-\frac{ 2 {\rm i} \pi}{\phi_0}\int_i^j \mathbf{A}\cdot\mathbf{dl}},
\label{eq:peierls}
\end{equation}
%
%
where $\phi_0$ is the flux quantum and $\mathbf{A}$ is a vector potential  associated to the magnetic field: $\mathbf{B}=\nabla \times \mathbf{A}$. For a given plaquette, we also introduce the magnetic flux $\phi$  in this plaquette and the reduced flux $f=\phi/\phi_0$.  

With the compactified regular $\{p,q\}$ graphs, it is simpler to consider dimensionless quantities. We must first ensure that faces  are coherently oriented, and then associate to each oriented edge a Peierls phase  such that their product around each face is a constant ${\rm e}^{2 {\rm i } \pi f}$. For the $\{p,q\}$ graphs, with $F$ identical faces, the reduced flux $f$ takes discrete allowed values $j/F$ with $j\in \mathbb{Z}$. In the $\mathcal{K}_p$ case, with two types of faces, the allowed  values of $f$ are slightly different, and given in Appendix~\ref{appendix:gauge}, which also describes in more detail the gauge construction. Plotting the eigenvalues versus $f$ leads to the Hofstadter butterflies presented below. 
%
%
\begin{figure*}[t]
\includegraphics[width=0.4\textwidth]{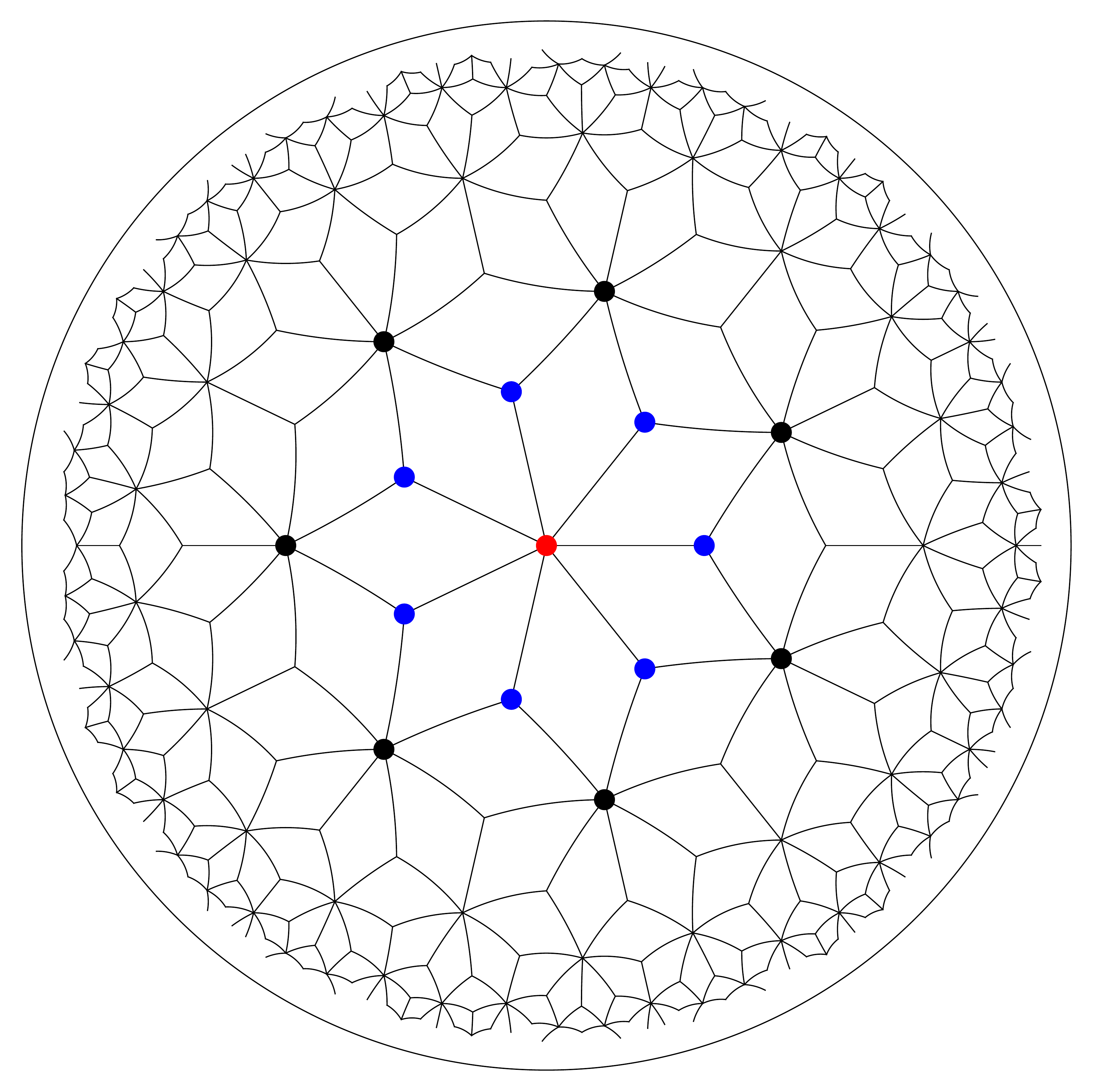} \qquad
\includegraphics[width=0.4\textwidth]{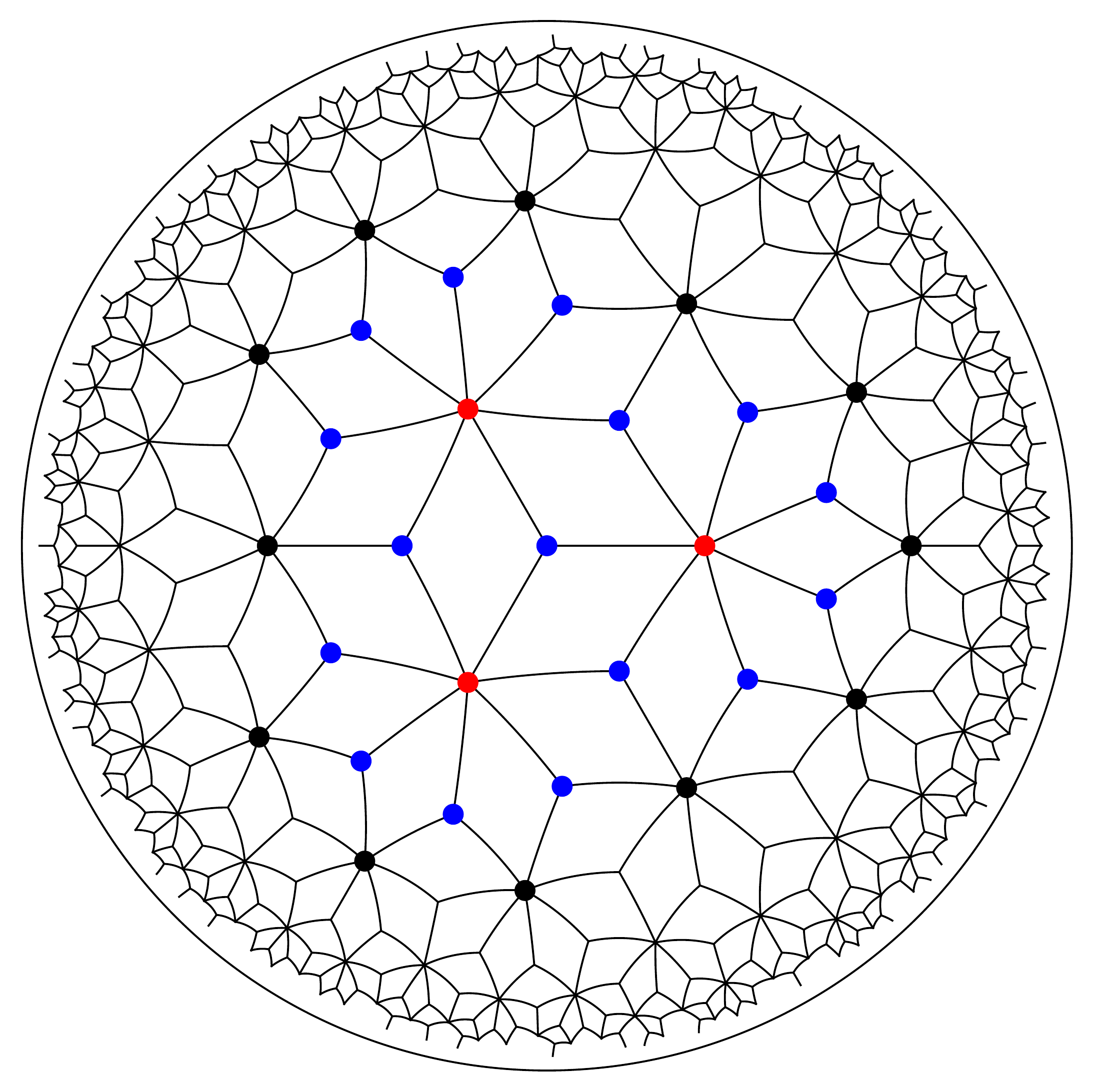}
\caption{Examples of Aharonov-Bohm cages for the $D_7$ tiling, which occur at $f=1/2$. Red (blue) disks are the sevenfold (threefold) coordinated sites inside the cage. Black disks correspond to the external boundary of the AB cages, namely, those sites where the destructive interference occurs at the critical flux; these sites are therefore not visited by a particle initially located at the central site, which remains trapped and  bounces back and forth among the red and blue sites. Left (right): AB cages associated with an initial sevenfold (threefold) coordinated site.}
\label{fig:cages}
\end{figure*}
%
%

%
%
\subsection{ Hyperbolic dice tilings and AB cages} 
\label{sec:ABcages}
%
%
The spectral properties of the Euclidean dice tiling $\mathcal{D}_6$ are well known (see Refs.~\cite{Horiguchi74,Sutherland86}) in the absence of magnetic field, and  when magnetic field is switched on (Ref.~\citep{Vidal98,Vidal01}). Since the $\mathcal{D}_p$ rhombic faces have equal area, the spectrum of $H$  is periodic with $f$, and we can restrict the study to $f \in [0,1]$. As for the above $\{p,q\}$ cases, the allowed values of $f_j$ are still of the form $j/F$, with $F$ the number of $\mathcal{D}_p$ faces. Furthermore, these tilings being bipartite, the energy spectrum is  symmetric with respect to zero.

Figure~\ref{fig:Butterfly_dice_main} shows the Hofstadter butterfly of several hyperbolic dice $\mathcal{D}_p$ tilings, with $p=7,...,11$, along with  the $p=6$ Euclidean case for comparison. The $\mathcal{D}_p$ sites' average coordination number (and therefore the overall butterfly width) is $p$ dependent. To compare the different butterflies, we have rescaled them by a factor of $\sqrt{3 p}$ in energy ($-\sqrt{3 p}$ being the Perron-Frobenius eigenvalue for $f=0$), leading to spectra in the range $[-1,+1]$.

As already discussed for other hyperbolic tilings~\cite{Stegmaier22}, namely, the $\{p,3\}$ and $\{3,p\}$ dual tilings, the hyperbolic butterflies are much less structured than the Euclidean ones. Nevertheless, they display some interesting features that we now describe. 

At vanishing flux we already notice a characteristic feature of hyperbolic tilings: the fact that the (Perron-Frobenius) lowest-energy state  is separated by a gap from the rest of the spectrum. The $\mathcal{D}_p$ tilings being bipartite, the same behavior occurs in the upper part of the spectrum. This gap remains finite at small fluxes but eventually vanishes.

But the main characteristic feature is clearly what occurs for $f=1/2$, where the spectrum reduces to three highly degenerate levels, at energy $\varepsilon=0, \pm \sqrt{p}$.
This is analogous to what has already been described in the original dice tiling ($p=6$)~\cite{Vidal98}. 

This peculiar situation is again described in terms of an Aharonov-Bohm cage: a particle initially located at any $\mathcal{D}_p$ site will remain trapped inside a small-size cage, whose size depends on whether the initial site is $p$-fold or threefold coordinated: for $p$-fold sites the cage consists of the initial site and the first shell with $p$ sites; for threefold sites, the cage consists of the initial site and the two neighboring shells. As recalled in the Introduction, this trapping  is due to an interference effect tuned by the magnetic field, that becomes completely destructive for $f=1/2$. Examples of cages are shown in Fig.~\ref{fig:cages}, for the $p=7$ case.

A nice way to characterize  cages, as was originally done in Ref.~\cite{Vidal98}, is by a Lanczos tridiagonalization of local clusters, using the recursion algorithm method~\cite{Haydock75}. The reduced Hamiltonian is that of a half chain (starting at the chosen initial site) with off-diagonal terms $b_j$ and diagonal terms $a_j$, the latter vanishing in the dice case due to the graph bipartiteness. The vanishing of one recursion coefficient $b_j$ implies that a particle initially localized at a given site will never escape from the neighboring shell associated with this vanishing coefficient, and will bounce back and forth within the cage. The $b_j$ coefficients are as follows:
%
%
\begin{itemize}

\item For a cluster centered on a $p$-fold coordinated site [see Fig.~\ref{fig:cages} (left)], the first two coefficients are
%
%
\begin{eqnarray}
b_1&=&-\sqrt{p}, \\
b_2&=& -2 \cos( \pi f),
\label{coeffp}
\end{eqnarray}
%
%
with $b_2$ vanishing at $f=1/2$, which leads to the AB cage phenomenon. This is directly related to the fact that the AB cage boundary is at the second-neighbor shell from the central site.

\item For a cluster centered on a threefold coordinated site [see Fig.~\ref{fig:cages} (right)], the first three coefficients are
%
%
\begin{eqnarray}
b_1&=& -\sqrt{3}, \\
b_2&=& -\sqrt{4 \cos^2(\pi f) + p - 3}, \\
b_3&=& -2 \cos (\pi f) \sqrt{\frac{p-4 \sin^2(2\pi f)}{4 \cos^2(\pi f)+p-3}} .
\end{eqnarray}
%
%
Note that it is now the third coefficient $b_3$ that vanishes at $f=1/2$, which is consistent with the fact that the cage has one additional layer in that case. 
\end{itemize}
%
%

A last interesting feature concerns the highly degenerate energy level at vanishing energy, which is present   for any magnetic flux, with spectral weight $(p-3)/(p+3)$. In the case of hyperbolic dice tilings $\mathcal{D}_p$, in contrast with the Euclidean $\mathcal{D}_6$, these vanishing energy levels are separated by a gap from the rest of the spectrum, for any $f$. We notice also that in some cases (such as $p=8$) some states leave the vanishing energy level when $f$ increases, a point that certainly  deserves further study. 
From the above values $b_1$ and  $b_2$, one understands that highly degenerate levels also occur for $f=1/2$ at $\varepsilon=\pm \sqrt{p}$, with spectral weight  equal to $3/(p+3)$ in both cases.

As a final remark, let us stress that, as already noted in the Euclidean case~\cite{Vidal01}, the butterfly of $\mathcal{D}_p$ tilings is related to the butterfly of the $\{3,p\}$ tiling. More precisely, to each eigenvalue $\varepsilon'(f)$  of the $\{3,p\}$ butterfly, there correspond two $\mathcal{D}_p$ butterfly eigenvalues: \mbox{ $\varepsilon_\pm(2f/3)=\pm \sqrt{ p-2 \cos( 2\pi f/3) \varepsilon'(f)}$}. Note that the highly degenerate zero-energy flat band of the $\mathcal{D}_p$ tilings is not captured by this mapping.

%
\begin{figure*}[ht]
\includegraphics[width=1.0\textwidth]{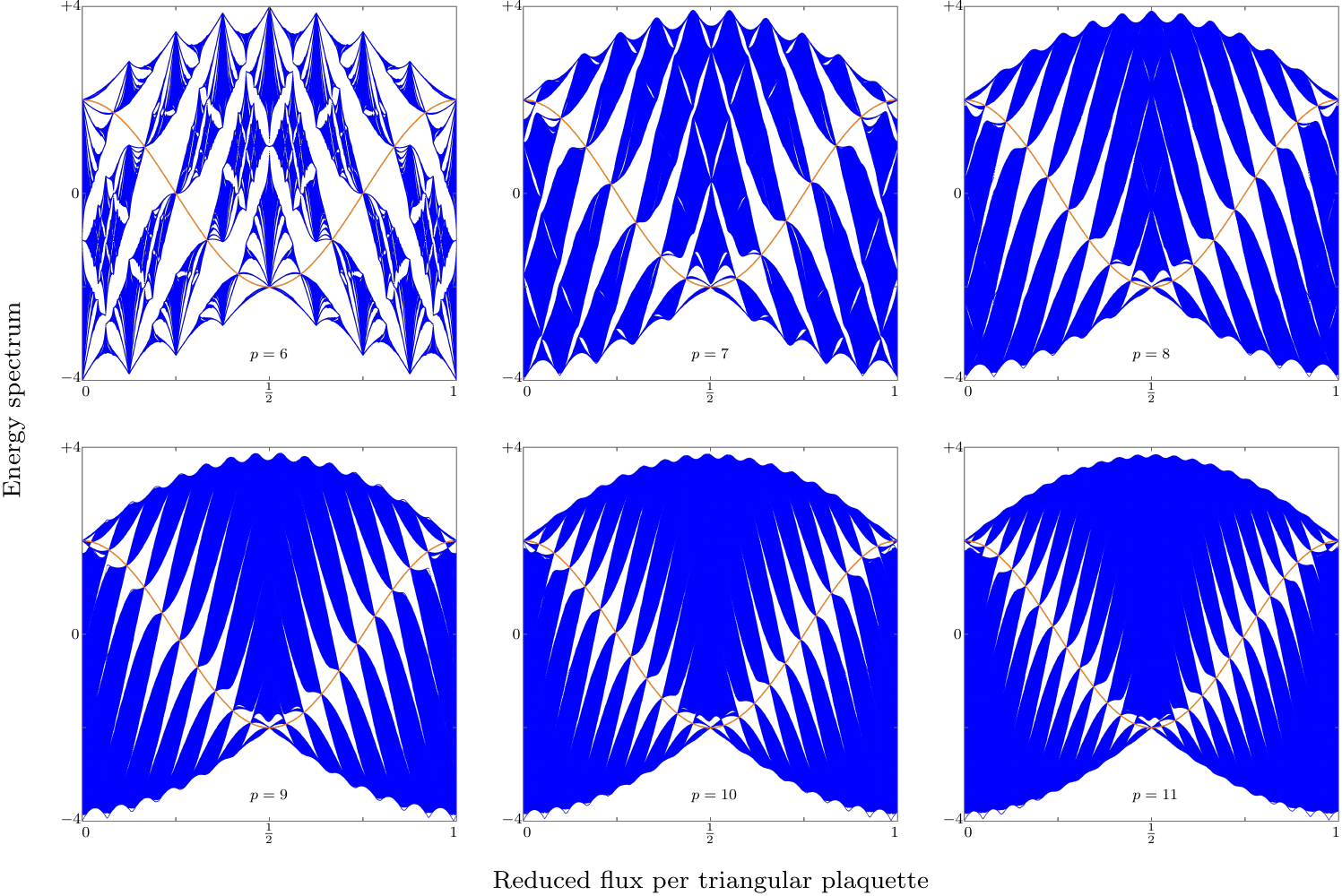}
\caption{Hofstadter butterflies for several hyperbolic kagome $\mathcal{K}_p$ tilings, displaying the spectral support versus magnetic flux. The different structures are characterized by the triplet ($p$, symmetry; $g$, surface genus; $V$, number of $\mathcal{K}_p$ vertices) which reads here: (6, 1, 3024), (7, 118, 4914), (8, 129, 3072), (9, 169, 3024), (10, 169, 2520),  (11, 231, 3036). All butterflies display an interesting sequence (increasing with $p$) of gaps and bands, the latter shrinking onto a curve with equation $\varepsilon (f)= 2\cos(2\pi f)$, drawn in orange.}
\label{fig:Butterfly_kagome_main}
\end{figure*}
%
%

%
%
\subsection{Hyperbolic kagome tilings} 
\label{sec:kagomebutterfly}
%
%

Let us now study the case of hyperbolic kagome tilings $\mathcal{K}_p$. As detailed in Appendix~\ref{appendix:gauge}, in order to get the Hamiltonian Peierls terms associated to a constant perpendicular magnetic flux (together with the characterization of allowed flux values), we must compute  the ratio $r_p$ of the $p$-gon area to the triangle area. In the Euclidean kagome case, the area ratio $r_6$ is exactly 6. In the hyperbolic case, the ratio $r_p$ reads: 
%
%
\begin{equation}
r_p=\frac {\pi (p - 2) - 
   2 p\arcsin \left( \frac {2\cos\left( \frac {\pi} {p} \right)}
{\sqrt {2\cos\left( \frac {2\pi} {p} \right) + 3}} \right)} {\pi - 
   6\arcsin \left( \frac {1} {\sqrt {2\cos\left( \frac {2\pi} {p} \right) + 3}} \right)},
\label{eq:area}
\end{equation}
which seems likely to be generically irrational. This would lead to nonperiodic butterflies except for the Euclidean case $p=6$.

Figure \ref{fig:Butterfly_kagome_main} shows  butterflies for several  hyperbolic kagome $\mathcal{K}_p$ tilings, with $p$ running from $7$ to $11$, along with the Euclidean case $p=6$  for comparison.

At vanishing flux, all hyperbolic kagome tiling butterflies display a highly degenerate higher energy level at $\varepsilon= 2$, which has been analyzed as a flat band in reciprocal space for the Euclidean $\mathcal{K}_6$~\cite{Xiao03}, but can as well (and maybe more simply) be associated to generic properties of line-graph adjacency  spectra~\cite{Biggs74}. Notice that these approaches have a sign difference since an adjacency matrix has $+1$ entries while $H$ has $-1$ hopping terms.

 The degeneracy found at $\varepsilon= 2$ depends on the bipartiteness of the graph. In the nonbipartite case, it is equal to $E-V$, where $E$ and $V$ are the number of edges and vertices of the original $\{p,3\}$ for which $\mathcal{K}_p$ is the line graph. In the bipartite case, there is an additional state, which has been recently attributed to a band-touching phenomenon~\cite{Maciejko22}. We give here a simple argument for this additional state, which holds for any line graph of a threefold coordinated graph $G$ (finite or infinite, ordered or disordered).  $G$ has a (nondegenerate) Perron-Frobenius eigenvalue $\varepsilon=-3$; in the bipartite case, it has an additional (nondegenerate)  opposite eigenvalue at $\varepsilon=+3$. For nonbipartite graphs, this state is absent and the upper part of the spectrum does not reach this value. Now, a general result for line-graph spectra~\cite{Biggs74}, adapted to the negative hopping term case, states that to any eigenvalue $\varepsilon$ in the $G$ spectrum there corresponds an eigenvalue $\varepsilon -1$ in its line-graph spectrum. As a consequence, limited to the bipartite case, an additional state is found at energy $+2$.

The  Perron-Frobenius state at vanishing flux and energy $\varepsilon=-4$ can be seen at the lower edge of the spectrum, separated from the rest of the spectrum by a gap that  eventually closes as the flux increases. Looking at the extremal part of the spectrum, one sees oscillations with flux, and even  a periodic reentrance of this gap. For a suitable gauge choice, one can show (see Appendix~\ref{appendix:perron}) that the associated eigenstate for vanishing flux (namely, a state with identical amplitudes on each site) is also an eigenstate at fluxes $f_j=j/(p/3+r_p)$ with energies $\varepsilon_j= -4 \cos(2\pi f_j/3)$. This peculiar behavior is already present for the original kagome tiling ($p=6$) butterfly but, as far as we know, was not yet noticed.

As can be seen in Fig.~\ref{fig:Butterfly_kagome_main}, $\mathcal{K}_p$ butterflies display quite interesting additional features:

%
%
\begin{itemize}

\item As $p$ increases, the number of gaps and bands increases, and the bands shrink, for specific flux values $f_j$, onto a curve with equation \mbox{$\varepsilon(f)=2 \cos(2 \pi f)$}.  
As explained in Appendix \ref{appendix:coscurvekagome}, the $f_j$ values vary with $p$ in a way that depends on the parity of $p$. For even $p$, $f_j=j/(p+r_p)$, while for odd $p$, $f_j=(j+1/2)/(p+r_p)$. Here again, this behavior is already present for the Euclidean $\mathcal{K}_6$. Notice that these specific flux values are expected for the infinite tiling in $\mathbf{H}^2$, but are generically not compatible with the flux quantization condition associated to PBCs (discussed in Appendix~\ref{appendix:gauge}). However, the latter values form a mesh that gets finer upon increasing the system size, and therefore gets very close to the above $f_j$'s.

\item Finally, the overall butterfly envelope seems to converge to an asymptotic shape as $p$ increases.
In order to study this shape, we consider  in Sec.~\ref{sec:cactus} the asymptotic $K_\infty$ tiling, which is nothing but the well-known triangular Husimi cactus, embedded in the hyperbolic plane~\cite{Mosseri82}.

\end{itemize}

%

%


\subsection{Hofstadter butterfly of the triangular Husimi cactus tiling}
\label{sec:cactus}
%
%

In this section, we compute the tight-binding spectrum under magnetic field for the triangular cactus tiling, which is the line graph of the trivalent Bethe lattice. At vanishing flux, the spectrum is well known and easily obtained~\cite{Thorpe81}. We provide here the extension for finite flux, which requires some changes in the original approach.

\begin{figure}[t]
\includegraphics[width=0.2\textwidth]{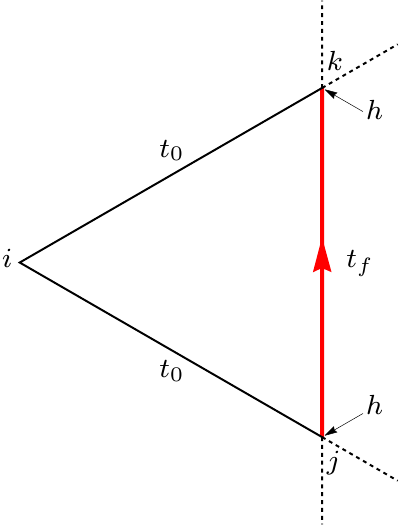}
\caption{Schematic of parameters used for the triangular Husimi cactus spectrum calculation. We focus on a cactus triangle (with sites $i, j$, and $k$) for which site $i$ is disconnected on the left from the rest of the tiling, and sites $j$ and $k$ are assigned a diagonal self-energy $h$ aiming to represent the effect of the rest of the tiling connected to these sites. A Peierls term $t_f$ is assigned to the edge connecting sites $j$ and $k$.}
\label{fig:cactusscheme}
\end{figure}

Let us first recall the computation in the vanishing flux case, whose main ingredients are shown in Fig.~\ref{fig:cactusscheme}. One considers an isolated triangular face of  the cactus tiling: one site (noted $i$) is cut from the rest of the tiling on one side, and  a (diagonal) self-energy $h$ (to be further determined in a self-consistent manner) is assigned to the two other sites of the triangle, noted $j$ and $k$, aiming to represent the effect of the rest of the tiling connected to these two sites. The Hamiltonian hopping term between $j$ and $k$ is  $t_f=t_0 \, {\rm e}^{2 {\rm i} \pi f}$, while the other two hopping terms are set to $t_0$. This choice ensures a reduced flux $f$ per triangle. Notice that here, in contrast with Eq.~(\ref{eq:hamiltonian}), we include the overall negative sign of the hopping term in the definition of $t_0$. The secular equations for an eigenstate with amplitudes $a_i, a_j$, and $a_k$ and eigenvalue $\varepsilon$ read:
\begin{eqnarray}
\varepsilon a_i&=& t_0 (a_j+a_k) +R, \\
(\varepsilon-h) a_j&=& t_0 a_i+\overline{t_f} a_k, \\
(\varepsilon-h) a_k&=& t_0 a_i+ t_f a_j,
\label{gaussbonnet}
\end{eqnarray}
where $\overline{t}$ denotes the complex conjugate of $t$ and $R$ is  the contribution that comes from the part that has been cut off at site $i$.

%
%
\begin{figure*}[ht]
\includegraphics[width=\textwidth]{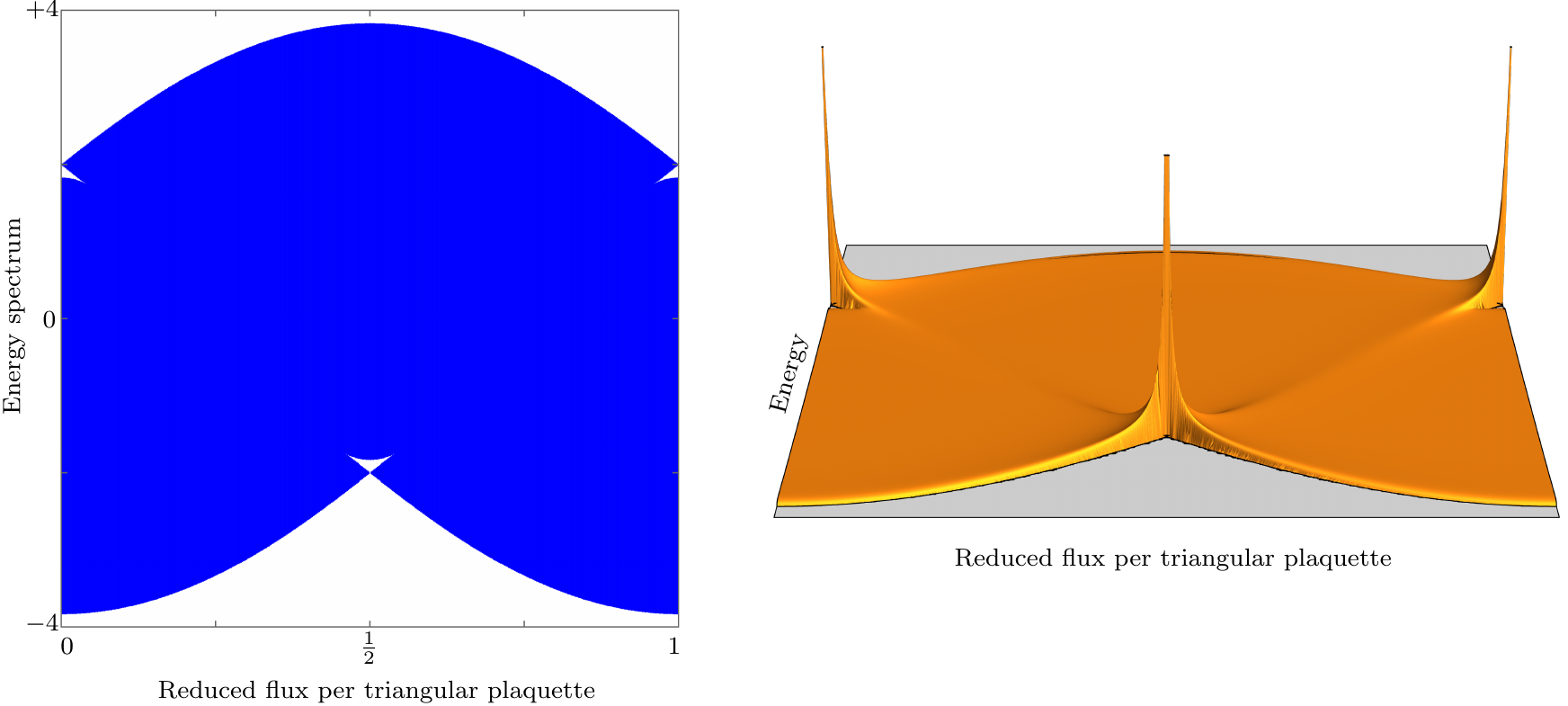}
\caption{  Left: Hofstadter butterfly  of the triangular Husimi  cactus (corresponding to the hyperbolic $\mathcal{K}_\infty$ tiling); its overall shape is the limiting shape for the hyperbolic $\mathcal{K}_p$ butterflies (see Fig.~\ref{fig:Butterfly_kagome_main}). At vanishing flux, the spectrum is made (for $t_0=-1$) of  a band in the range $[-1-2\sqrt{2},-1+2\sqrt{2}]$ with spectral weight $2/3$, a flat band at $\varepsilon=2$ with spectral weight $1/3$, and an isolated nondegenerate Perron-Frobenius state at $\varepsilon=-4$ (not shown here). Small gaps are visible near the flux values $0$ and $1/2$. Right: Density of states versus magnetic flux, showing divergences close to the gaps and a local maximum along the $\varepsilon(f) =2 \cos (2 \pi f)$ line.}
\label{fig:cactusspectrum}
\end{figure*}
%
%

By setting $f=0$, we can repeat the analysis already given in Ref.~\cite{Thorpe81}. In this case, $a_j$ and $a_k$ can be eliminated in the above equations, leading to

%
%
\beq
a_i \left(\varepsilon-\frac{2t_0^2}{\varepsilon-h-t_0}\right)=R.
\eeq 
%
%
The self-energy $h$ therefore satisfies a quadratic equation derived from the self-consistent condition:
%
%
\beq
h=\frac{2t_0^2}{\varepsilon-h-t_0}.
\eeq 
%
%
From the  local Green's function of the full cactus tiling at site $i$, $G_{ii}=1/(\varepsilon-2h)$, we obtain, for the density of states,

%
%
\beq
n(\varepsilon)=-{\rm Im} \,( G_{ii}/\pi)= \frac{1}{\pi}\frac{\sqrt{8t_0^2-(\varepsilon-t_0) ^2}}{9t_0^2-(\varepsilon-t_0)^2}.
\eeq 
%
%

This expression only captures $2/3$ of the cactus spectrum. Indeed, one must add the highly degenerate level, already discussed above, at energy $\varepsilon=-2 t_0$, characteristic of the line graph of a trivalent graph. Also missing is the (isolated) nondegenerate Perron-Frobenius state at \mbox{$\varepsilon= 4 t_0$}.

We now consider the case of a nonvanishing flux. The Peierls phase differentiates sites $j$ and $k$, which requires additional manipulations to get the flux-dependent self-energy $h_f$. The latter now satisfies a cubic equation derived from the self-consistent condition
%
%

\beq
h_f=\frac{2 t_0^2[\varepsilon-h_f+t_0\cos(2 \pi f)]}{(\varepsilon-h_f)^2-t_0^2}.
\eeq 
%
%
Looking for solutions to this cubic equation with nonvanishing imaginary part  leads to the triangular Husimi cactus Hofstadter butterfly shown in Fig.~\ref{fig:cactusspectrum} (left). We can furthermore derive, as above, the density of states $n(\varepsilon,f)$, which is plotted  in Fig.~\ref{fig:cactusspectrum} (right).

The sequence of fluxes where the spectrum pinches, which was already increasingly tight as $p$ increases, is no longer visible, nor are the associated gaps. However, regarding the density of states, we see that this set of highly degenerate levels leads to  a maximum of the continuous density of states along the $\varepsilon(f) = 2 \cos(2 \pi f)$ curve, recalling what happens when discrete levels enter a continuous band and are turned into resonating levels.

In conclusion, we have shown that the triangular cactus butterfly can be computed exactly, showing (i) an asymptotic envelope  close to what could be expected from Fig. \ref{fig:Butterfly_kagome_main}, (ii) a vanishing of the gap sequence, and (iii) in place of the discrete pinched spectrum with highly degenerate levels,  a maximum of the density of states along the $ 2 \cos(2 \pi f)$ curve.

%
%
\section{Gap labeling}
\label{sec:gaplabel}
%
%

For Euclidean lattices, gaps of the Hofstadter butterfly can be labeled by two integers, the Chern number and another integer which is reminiscent of the band structure~\cite{Wannier78}. 
The goal of this section is to propose a similar labeling for hyperbolic dice and kagome lattices. To this aim, we first recall the essential steps of the Euclidean case and then compute the Wannier diagrams for hyperbolic tilings.

The Hall conductivity $\sigma_{\rm H}$ at energy $\varepsilon$ inside a gap is given by the Widom-St$\check{\rm{r}}$eda formula~\cite{Streda82,Widom82}:  
%
%
\beq
\sigma_{\rm H} =  \frac{e}{\mathcal{A}_{\rm tot}}  \frac{\partial N(\varepsilon,B)}{\partial  B}= -\frac{e^2}{h} \nu,
\label{eq:Hall}
\eeq 
%
%
where $\nu$ is a topologically invariant integer called the Chern number, $\mathcal{A}_{\rm tot}$ is the total area of the system, and $N (\varepsilon,B)$ is the integrated density of states. This indicates that the total number of states below the gap is given by:
%
%
\beq
N(\varepsilon,B) = \nu B \, \mathcal{A}_{\rm tot}+ \lambda ,
\label{eq:N_of_f}
\eeq 
%
%
where we set the flux quantum $\phi_0=h/|e|=1$. In the presence of periodic boundary conditions, the total flux through the surface $B \, \mathcal{A}_{\rm tot}$ must be an integer (in units of $\phi_0$) as well as $\lambda$.

For any Euclidean lattice with $N_{\rm u}$ unit cells~\cite{Wannier78,Streda82,Thouless82}, the integrated density of states per unit cell can be decomposed as: 

%
%
\begin{equation}
\mathcal{N}(\varepsilon,B)=\frac{N(\varepsilon,B)}{N_{\rm u}} = \nu f + \mu,
\label{eq:IDOS}
\end{equation}
%
%
where $\mu$ is an integer, and $f=B \, \mathcal{A}_{\rm tot}/N_{\rm u}$ is  the reduced flux per unit cell. Thus, any gap can be labeled by $(\nu,\mu)\in \mathbb{Z}^2$.

 As discussed in Sec.~\ref{sec:PBC}, the description of structures of increasing sizes is very different in the Euclidean and the hyperbolic cases. In the former case, all structures are defined on a $g=1$ torus, and  the total number of sites is proportional to $N_{\rm u}$, which
leads to the above normalization procedure.

The hyperbolic case is quite different. Larger and larger hyperbolic tiling patches are folded onto tori of increasing genus.
All interesting parameters scale with the quantity $(g-1)$: this is true for the number of sites, as shown in Sec.~\ref{sec:PBC}; this also applies to $g$-holed tori, whose area, from the Gauss-Bonnet relation, reads $4\pi (g-1)$. So, clearly, one expects that $(g-1)$ plays the role of $N_{\rm u}$ in order to normalize the density of states in the hyperbolic case.

\begin{figure}[t]
\includegraphics[width=1\columnwidth]{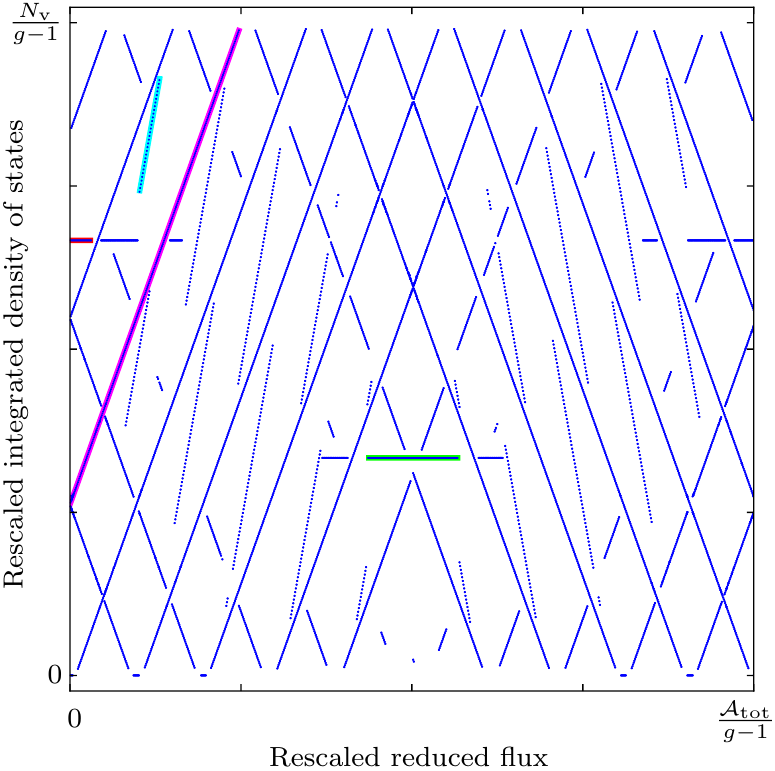}
\caption{ Wannier diagram of the $\mathcal{K}_7$ tiling with $N_{\rm v}=4914$ vertices ($g=118$). For a given value of the reduced flux $f$, each blue dot indicates the integrated density of states,  $\mathcal{N}$, in a given gap, both quantities being normalized by $g-1$. A gap is identified as a difference between two consecutive energy levels larger than 
$\delta=0.01$. Each gap (blue line) can be indexed by two integers $(\nu,\mu)$ according to Eq.~(\ref{eq:IDOS}), with the above prescribed normalization. Here, we label four gaps indexed by $(0,28)$ (red), $(0,14)$ (green),  $(1,11)$ (magenta),  and $(2,6)$ (cyan). With this value of $\delta$, we only observe gaps with $\nu=0, \pm 1,\pm 2$ but we cannot exclude the existence of smaller gaps below this threshold.
}
\label{fig:Wannier}
\end{figure}

 To clearly identify and label these gaps, the simplest method is to compute the Wannier diagrams (see Ref.~\cite{Fuchs16} for a similar approach).  These diagrams are obtained by plotting $\mathcal{N}=N/(g-1)$ as a function of $f$  for each gap greater than a given threshold $\delta$ that we select appropriately. Ideally, one should consider the limit where $\delta$ vanishes but since we deal with finite-size systems with a few thousands sites,  we choose $\delta$ to be much larger than the typical level spacing. After checking that each gap identified is stable while increasing the system size, it is straightforward to extract $(\nu,\mu)$. To illustrate this methodology, we present in Fig.~\ref{fig:Wannier} the Wannier diagram associated with the Hofstadter butterfly of the  $\mathcal{K}_7$ tiling. Note that an alternative approach has been recently used to compute Chern numbers in hyperbolic tilings~\cite{Liu22}. 
%
%
%
%

In the Euclidean case ($p=6$), all possible Chern numbers are found in the Hofstadter butterflies of $\mathcal{D}_6$ and $\mathcal{K}_6$  as is always the case for periodic lattices~\cite{Wannier78,Claro79} (see also Refs.~\cite{Osadchy01,Agazzi14,Avron14}). As can be seen in Figs.~\ref{fig:Butterfly_dice_main} and \ref{fig:Butterfly_kagome_main}, the gap structure of $\mathcal{D}_p$ becomes simpler and simpler when $p$ increases. In the large-$p$ limit, the Hofstadter butterfly of the $\mathcal{D}_p$ tiling only contains the two large central gaps below and above \mbox{$\varepsilon=0$} for which $\nu=0$. By contrast, for $\mathcal{K}_p$, one gets an increasing number of gaps with $\nu=\pm 1$ together with a few gaps with $\nu=0$.

%
%
\section{Conclusion}
\label{sec:ccl}
%
%

As compared to the case of Euclidean two-dimensional tight-binding systems under transverse magnetic field, hyperbolic tilings display much less structured  Hofstadter butterflies with, for instance, no evidence of self-similarity as displayed in the former case~\cite{Stegmaier22}. Nevertheless, we have shown that several interesting features could be found in the tilings considered in this work. In summary:

(i) We considered an infinite sequence of hyperbolic dice tilings $\mathcal{D}_p$, with alternating threefold and $p$-fold coordinated sites and identical rhombus tiles. All these tilings show the AB caging effect for a critical flux (while very few cases are known for Euclidean tilings). The spectrum pinches onto three highly degenerate energy levels at $\varepsilon=0,\,\pm \sqrt{p}$. The vanishing energy eigenvalue is present for any flux value, and is separated by gaps from the rest of the spectrum.

(ii)  We have also considered an infinite sequence of  hyperbolic kagome tilings $\mathcal{K}_p$, dual to the previous  $\mathcal{D}_p$, with fourfold coordinated sites and triangular and regular $p$-gonal faces. Their butterflies display gaps whose number grows (and width decreases) with $p$. We have also shown that these gaps close at discrete highly degenerate energies where the  spectrum pinches  along a simple cosine curve. Also interesting is a second discrete set of nondegenerate levels organized along another cosine curve whose period is three times larger. Notice that these two features are already present in the Euclidean kagome tiling and were not, to our best knowledge, already described. Interestingly, the hyperbolic $\mathcal{K}_p$ tilings generically present an apparently irrational value for the ratio of their regular $p$-gon to triangle areas, breaking the periodicity of the butterfly with respect to the magnetic field.

(iii) We have analyzed the limiting case of $\mathcal{K}_\infty$ which is nothing but the known triangular Husimi triangular cactus structure (the line graph of the threefold coordinated Bethe lattice). We extend to nonvanishing flux, the standard method to compute analytically the spectrum at zero flux. This amounts to computing a flux-dependent self-energy from the roots  of a  cubic polynomial (which is quadratic for vanishing flux). The corresponding butterfly envelope compares well to that of the large-$p$ cases. In particular, gaps, whose width decreases with $p$, disappear, and  the discrete set of  highly degenerate states along a cosine curve (discussed above) is transformed into a local maximum along the same curve in the $\mathcal{K}_\infty$ flux-dependent density of states.  

(iv)   Finally, we also proposed a gap labeling for hyperbolic tilings and used a direct procedure to compute the Chern number via the Wannier diagrams. Interestingly, only gaps with small Chern numbers ($|\nu|\leqslant 2$) have been found, although we cannot rule out the existence of smaller gaps with larger $\nu$. \\

An important step in getting a well-defined spectrum is to apply PBCs to get rid of edge states. For that purpose, we proceeded with the present regular $\mathcal{D}_p$ and $\mathcal{K}_p$ tilings as was previously done in Ref.~\cite{Stegmaier22} for regular $\{p,q\}$ tilings, mapping larger and larger tiling patches onto $g$-holed tori. The approach to infinite-size tilings is much more complex than in the Euclidean case, dealing with tori having increasing numbers of holes, and noncontractible loops whose size increases quite slowly.

As recalled in the Introduction, several experimental implementations have been proposed and/or realized in the context of the AB caging effect for Euclidean tilings; the case of finite patches of hyperbolic tilings  has also been considered at vanishing magnetic flux. We can therefore reasonably expect that some of the properties described here can be realized in real systems. 

{\em Note added:}  Recently, a related study was posted~\cite{Maciejko22}, also dealing with hyperbolic dice and kagome tiling spectra, but in the absence of magnetic field. \\
%
%
\acknowledgements
%
%

We thank J.-N. Fuchs and J. Maciejko for fruitful discussions.

\appendix

%
%
\section {Hurwitz tilings}
\label{appendix:hurwitz}
%
%

The dual $\{7,3\}$ and $\{3,7\}$ tilings play a special role in the theory of hyperbolic geometry, with connections to other fields of mathematics. Of particular interest is the set of highly symmetrical finite  $\{7,3\}$ patches, defined on  $g$-holed tori, known as Hurwitz tilings.

Hurwitz's celebrated theorem states that the orientation-preserving (automorphism) group acting on a $g$-holed surface $S$  has maximal order $84 (g-1)$. The existence of such a bound can be understood by looking at the covering space of these compact surfaces, namely, the hyperbolic plane $\mathbf{H}^2$. The discrete group which acts on the compact surface has an associated polygonal fundamental region  which covers the surface under the group action. The order of the group is therefore the number of such regions covering $S$, which equals the ratio of the total area of $S$ to that of the fundamental region. 

Going to regular  $\{p,q\}$ tilings on $\mathbf{H}^2$ (with constant Gaussian curvature $\kappa=-1$), one can show that the smallest fundamental region is twice that of the $\{7,3\}$ group, with area $\pi /21$. From the Gauss-Bonnet theorem, we know that $S$ has area $4\pi (g-1)$, which gives $4\pi (g-1)/(\pi/21)= 84 (g-1)$ as the group order, thus reaching the Hurwitz bound.

The smallest Hurwitz surface is the celebrated genus-$3$ Klein quartic~\cite{Klein1878}. The first Hurwitz surfaces (with well-defined finite groups) are found to exist on tori of genus  3, 7, 14, 17, 118, 129, 146, 385, 411, 474,... They therefore provide a rich sequence to follow in order to study larger and larger hyperbolic tiling patches with PBCs.
%
%
\begin{figure*}[t]
 \includegraphics[width=\textwidth]{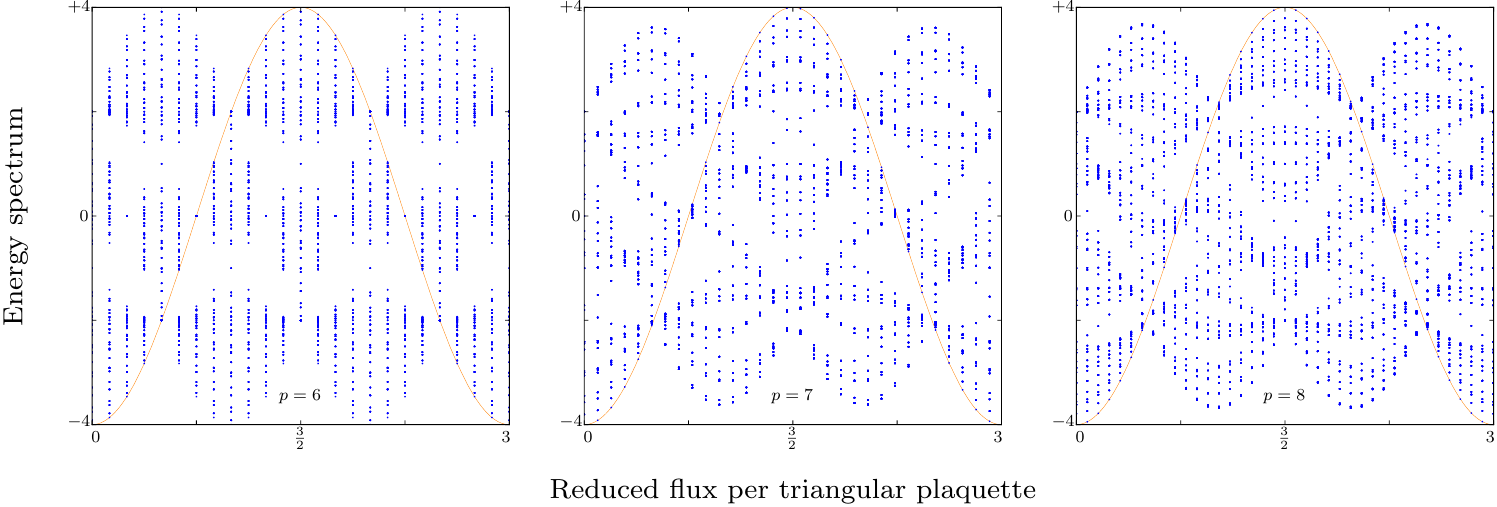}
\caption{ Illustration, for $\mathcal{K}_6,\mathcal{K}_7$, and $\mathcal{K}_8$ butterflies, of a discrete set of states at energy $\varepsilon_j=-4 \cos(2 \pi f_j/3)$ for $f_j=j/(p/3+r_p)$.  With the chosen gauge, the eigenstates are equal-amplitude (Perron-Frobenius-like) eigenstates.}
\label{fig:kagome_perron}
\end{figure*}
%
%

%
%
\section { Gauge construction}
\label{appendix:gauge}
%
%
In principle, computing the Peierls phase term in Eq.(\ref{eq:peierls}) requires choosing a vector potential  $\mathbf{A}$ and performing line integrals along the edges. This is quite easy for Euclidean tilings with explicit site coordinates and vector potential. In the present case, we work with tilings given as graphs (collection of vertices and edges) on genus-$g$ tori, defined by their adjacency matrix. We must therefore proceed differently.

In a first step, we must define a coherent orientation for the tiling faces, which reflects the orientability of the underlying genus-$g$  tori. This amounts to defining, for each face, local  edge orientations to circumnavigate the face, which must be opposite for the two adjacent faces sharing the given edge. A way to do this is to use a spanning tree of the dual tiling: first define an (arbitrary) orientation for the face associated with the root vertex, and then propagate the face orientation along the spanning tree. The orientability of the genus-$g$ tori guarantees that this procedure can be completed.

Once the coherent face orientation is defined,  the Peierls phase derivation can be addressed. We aim to compute the tight-binding spectrum in the presence of a constant perpendicular magnetic field. Whenever the faces have constant area, as in the hyperbolic dice tilings, this translates to a simple condition of equal product of (vector potential  dependent) Peierls terms around each face (see, for instance, Ref.~\cite{Avishai08}). A simple way to proceed is to use the tiling spanning tree and let the vector potential vanish on its edges. The condition for constant product of the hopping term around a face translates into a condition of constant sum of discrete vector potential terms $\mathbf{A}_{i,j}$, where ${i,j}$ denotes the edge between site $i$ and $j$. This leads to a set of coupled linear equations with (remaining) variables $\mathbf{A}_{i,j}$ which is carried for all faces except a last one. The constraint for this last face (up to a $2 \pi$ modulo operation in the product of Peierls terms), leads to a discrete set of allowed flux values. 

 When all faces have equal area (as in the $\mathcal{D}_p$ case), the Hofstadter butterfly is periodic with the reduced flux $f$, and the number of allowed values per period equals the ratio of the total area over a single face area, i.e., the number of faces. Now, for $\mathcal{K}_p$ tilings, the situation is different since one has $F_3$ triangles and $F_p$ $p$-gons, whose  area ratio $r_p$ is likely to be irrational. Hence, the butterfly lacks periodicity and the allowed flux values (counted with respect to the flux in a triangle)  read \mbox{$f_j=j/(F_3+r_p F_p)$}.

Notice that the above procedure lists fewer constraints than the number of available variables, leading to a set of 2$g$ free parameters which can be analyzed in terms of additional constraints on the noncontractible loops of the genus-$g$ surface. The authors of Ref.~\cite{Stegmaier22} perform averages over these additional fluxes, while we choose here to set these free parameters to zero. A discussion about the respective merit of these two choices is left for further discussions.

%
%
\begin{figure*}[t]
\includegraphics[width=0.45\textwidth]{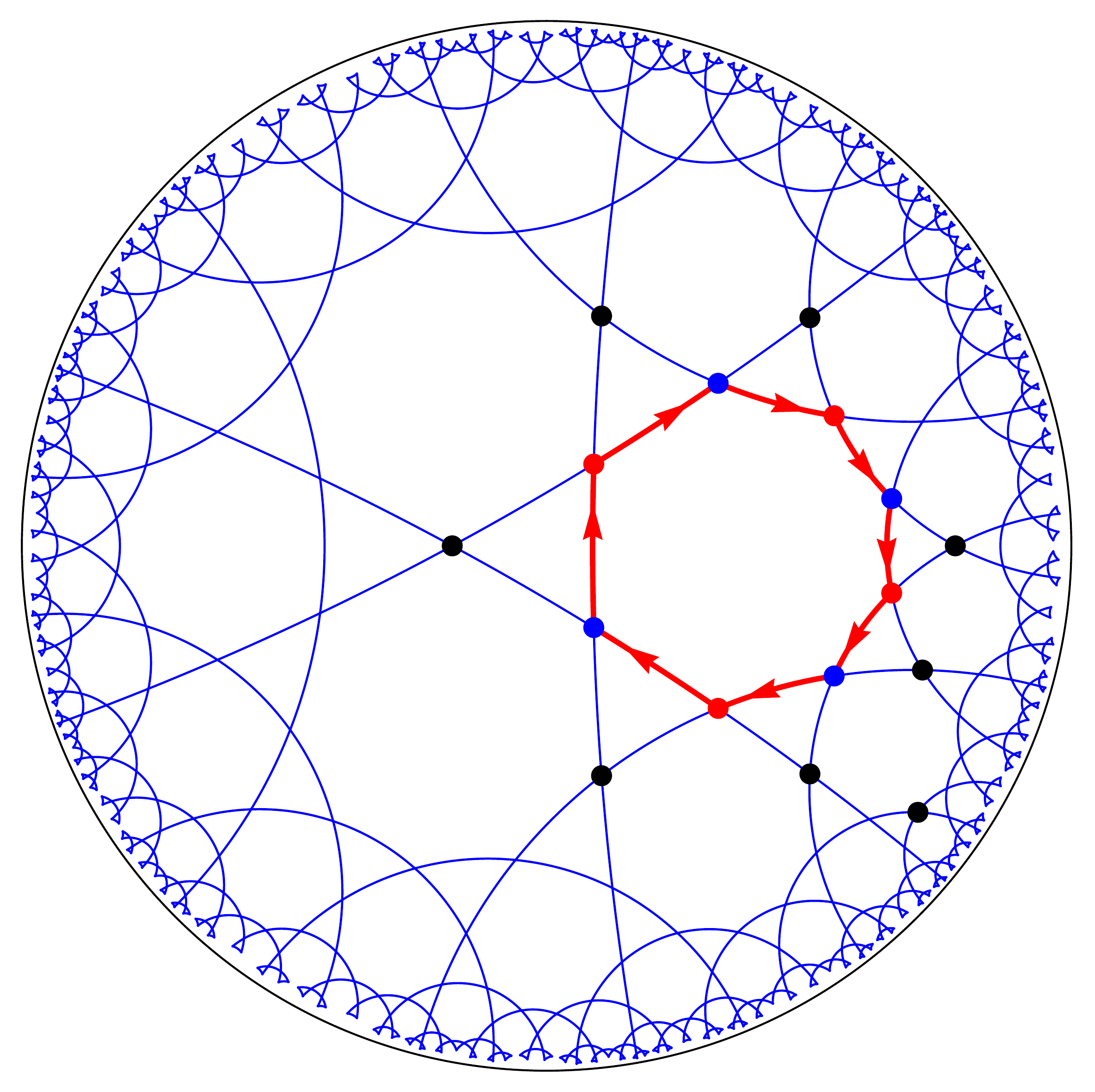}
\includegraphics[width=0.45\textwidth]{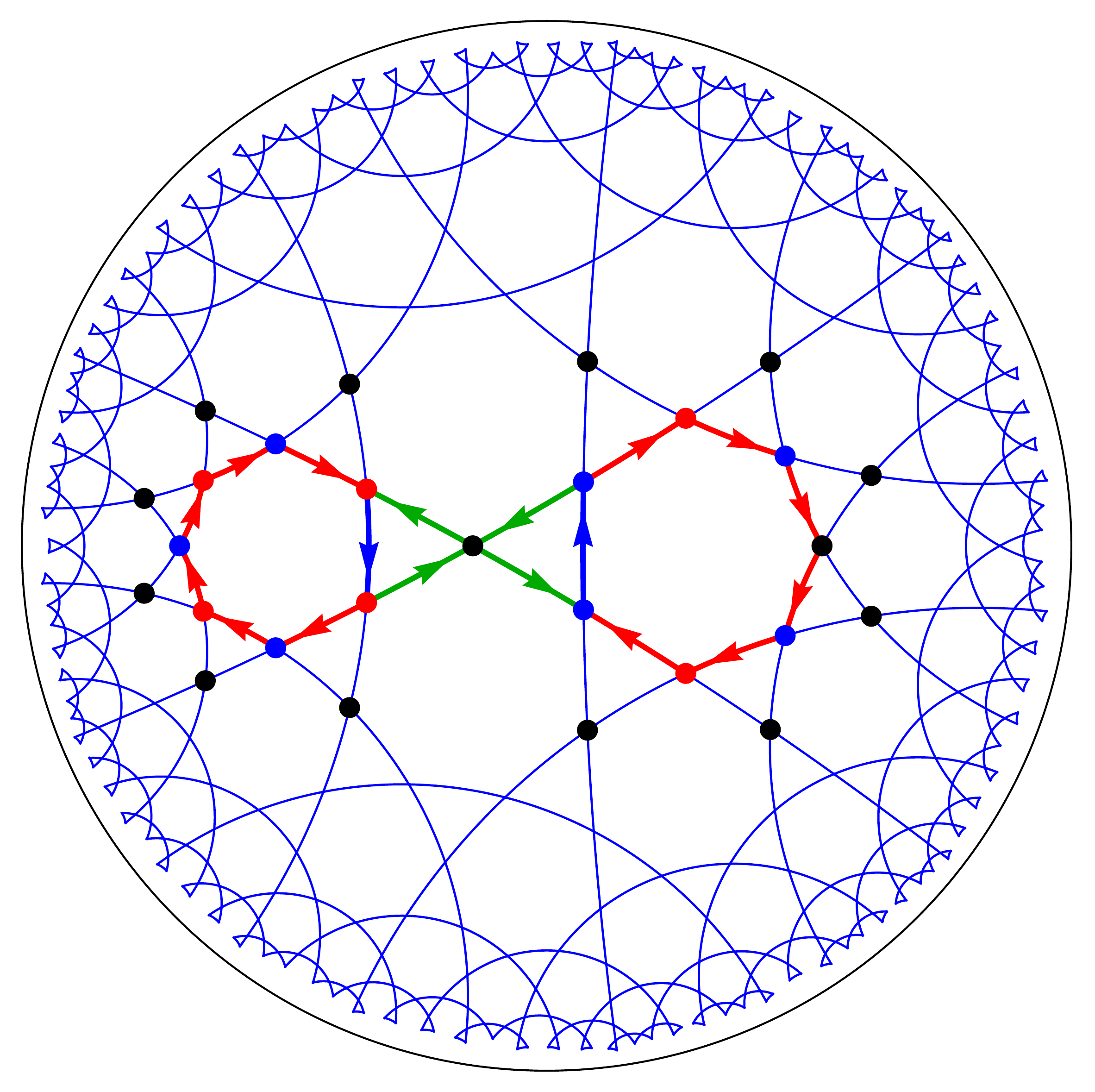}      

\caption{Local gauge for kagome tilings used to display the highly degenerate states at energy $\varepsilon=2 \cos(2 \pi f)$ for selected flux values. We use the following arrow color convention for the gauge hopping terms: (i) hopping term $t_0=-1$ (no arrow), (ii) term $t_f=-{\rm e}^{2 {\rm i}\pi f}$ (red arrow), (iii) term $-t_f$ (blue arrow), and (iv) term $ t_f={\rm e}^{ {\rm i} \pi/2}$ (green arrow).  Left, the even-$p$ case; right, the odd-$p$ case. The (unnormalized) eigenstates have alternating amplitudes $+1$ (red dots) and $-1$ (blue dots), and zero elsewhere. Black disks denote those sites with zero amplitude that immediately bound the localized eigenstate. The eigenvalue $\varepsilon(f)=2 \cos (2 \pi f)$ for such states can be directly verified.}
\label{fig:kagome_2cos}
\end{figure*}
%
%
\section { Some details about the tilings}
\label{appendix:datatilings}

For our numerical computation, we used two different data sets of adjacency matrices associated with symmetric finite patches of $\{p,3\}$ tilings defined on $g$-holed tori. One must then  operate a decoration of these graphs to produce the adjacency matrices of the associated  $\mathcal{D}_p$ and $\mathcal{K}_p$ graphs. In this work, we have named the studied structures by a triplet $(p, g, V)$, with   $p$ the symmetry order,  $g$ the surface genus, and $V$ the number of $\mathcal{D}_p$ or $\mathcal{K}_p$ sites. 

For $p=7$, we used a construction of the genus-$118$ Hurwitz tiling, generated by one of  us (R.V.), and available upon request.

For the other values of $p$, we used data given by Conder~\cite{Conder06}, who made accessible the adjacency matrices for 3-valent symmetric graphs up to $2048$ vertices. These data names use two numbers: the number of sites, followed by a ``type'' number. Table \ref{tab:conder} indicates which data in this set have been used in the present study.

\begin{table}[h]
\center
\begin{tabular}{|c | l |}
\hline
$p$ & File name \\
\hline
6 & C2016.2\\
\hline
8 & C2048.23\\
\hline
9 & C2016.6\\
\hline
10 & C1680.2\\
\hline
11 & C2024.2\\
\hline
\end{tabular}
\caption{List of the files from Ref.~\cite{Conder06} used in this work.}
\label{tab:conder}
\end{table}
%

%
\section {States with $\cos (2 \pi f/3)$ oscillations in the kagome butterfly }
\label{appendix:perron}
%
%

In this Appendix, we explain the occurrence of a simple sequence of eigenvalues at $\varepsilon_j = -4 \cos(2 \pi f_j/3)$  which are found for any $p$ value and for the discrete set of fluxes $f_j=j/(p/3+r_p)$. They are shown in Fig.~\ref{fig:kagome_perron} for \mbox{$p=6,7$, and $8$}. We consider the simple following  gauge, which can be defined on any $\mathcal{K}_p$ tiling: as done everywhere here, we only define the gauge by specifying the Peierls terms on the graph. In $\mathcal{K}_p$ tilings each edge is shared by a (unit area) triangle and a $p$-gon, and we can assign the edges an orientation corresponding to the same (say, anticlock wise)  orientation for all the triangles and the opposite orientation on the $p$-gons. The proposed gauge is now set by attaching a Peierls term ${\rm e}^{2 {\rm i} \pi f/3}$ to each oriented edge. As a result, their product around a triangle corresponds to having a flux $f$ threading the triangles. Now to achieve a constant flux through the whole tiling, we have the following constraint due to the $p$-gons:
${\rm e}^{-2 {\rm i } \pi p f/3}={\rm e}^{ 2 {\rm i } \pi f r_p}$ implying that \mbox{$f_j=j/(p/3+r_p)$} with $j\in \mathbb{Z}$. Recalling that for kagome tilings one has $p F_p= 3 F_3$, one verifies that these values of $f_j$ are compatible with those given in Appendix~\ref{appendix:gauge}.

With this discrete gauge, each $\mathcal{K}_p$ site has two entering and two outgoing arrows, carrying therefore ${\rm e}^{\pm 2 {\rm i } \pi f_j/3}$. As a result, a state with equal amplitude (say, $+1$ in unnormalized form) is an eigenstate with eigenvalue \mbox{$\varepsilon_j= -4 \cos(2 \pi f_j/3)$}. At vanishing flux, this corresponds to the standard Perron-Frobenius state occurring at the spectral lower edge.

Notice finally that the hyperbolic $\mathcal{K}_p$  butterflies seem to display some regular patterns when analyzed at the discrete set $f_j$. Their analysis is left for future work.

\section {Spectral pinched patterns in kagome tilings}
\label{appendix:coscurvekagome}

As discussed above, and visible in Fig.~\ref{fig:Butterfly_kagome_main}, $\mathcal{K}_p$ butterflies display an interesting sequence of gaps and bands, the latter shrinking, for selected $f$ values, onto a curve with equation $\varepsilon(f)=2\cos(2\pi f)$. We analyze here this question  in detail. 

Let us first recall that at vanishing flux, the kagome spectrum   shows a nondispersive flat band at energy \mbox{$\varepsilon=2 $} (see Ref.~\citep{Xiao03}). This is in fact related to a general result for so-called line-graph spectra, as discussed in Ref.~\cite{Biggs74}. Note that in the line-graph case, the spectrum is that of the adjacency matrix (with entries $+1$), leading to  a flat band at the eigenvalue $-2$, instead of $+2$ with our tight-binding Hamiltonian given in Eq. (\ref{eq:hamiltonian}).

The occurrence of the highly degenerate state at energy $\varepsilon=2$ has been discussed in Sec.~\ref{sec:kagomebutterfly}. Under a magnetic field, we find  that this property holds for a discrete set of fluxes, for other values of energy. This property is already present for the Euclidean kagome butterfly~\cite{Xiao03}, although apparently unnoticed. 

We show now that a local gauge can be defined for which  simple confined eigenstates are proved to exist.

$\mathcal{K}_p$ tilings are made of triangles and $p$-gons. Consider an isolated $p$-gon and its $p$ neighboring triangles, as shown in Fig. \ref{fig:kagome_2cos}. Fluxes are counted with respect to area: a flux $f$ in a triangle is ensured by a Peierls term  $t_f=-{\rm e}^{2 {\rm i} \pi f}$ on the edge shared by the $p$-gon and its $p$ neighboring triangles.
We separate the cases for even and odd $p$. 

{\em $p$ even.}
The local proposed gauge is shown in Fig. \ref{fig:kagome_2cos} (left). The product of hopping terms along the $p$-gon, in reversed orientation as compared to the triangle, leads to a quantity $\overline{t_f}^p={\rm e}^{-2 {\rm i} \pi p f}$, which should equal ${\rm e}^{2 {\rm i} \pi r_p f}$ to correspond to a uniform transverse magnetic field (recall that $r_p$ is the ratio of the $p$-gon to the triangle areas). This leads to a discrete set of $f$ values \mbox{$f_j=j/(p+r_p)$} indexed by an integer $j$.
Now, for those flux values $f_j$, one can exhibit a confined state with energy $\varepsilon_j=2\cos(2\pi f_j)$, described here in an unnormalized form: it has alternating amplitudes of $+1$ (say blue dots) and $-1$ values (red dots) at the vertices of the selected $p$-gon sites, and, on each of the neighboring triangles, vanishing amplitude at the remaining, distal site. Such a confined state is an eigenstate with the expected eigenvalue. Being isolated, it can be repeated with high degeneracy.

{\em $p$ odd.}
In this case, the local gauge construction is more complex (see Fig.~\ref{fig:kagome_2cos}, right); it involves two $p$-gons with some hopping term equal to ${\rm e}^{{\rm i} \pi/2}$. The selected fluxes now read $f_j=(j+1/2)/(p+r_p)$. The eigenstate at energy $\varepsilon_j=2\cos (2\pi f_j)$ would need an alternation of $+1$ and $-1$ amplitudes, which is impossible due to the odd value of $p$; this leads to one pair of neighboring sites sharing the same sign. This is compensated for by one Peierls term with opposite sign. Notice that one gets another eigenstate with the same energy by flipping the amplitude signs on one polygon, and that this eigenstate is orthogonal to the first one.

%

\end{document}